# Laser-induced Precession of Magnetization in GaMnAs

Eva Rozkotová[1], Petr Němec[1], Daniel Sprinzl[1], Petra Horodyská[1], František Trojánek[1], Petr Malý[1], Vít Novák[2], Kamil Olejník[2], Miroslav Cukr[2], and Tomáš Jungwirth[2]

[1] Faculty of Mathematics and Physics, Charles University in Prague, Prague, 121 16, Czech Republic
[2] Institute of Physics ASCR  v.v.i., Cukrovarnická 10, Prague, 162 53, Czech Republic

**We report on the photo-induced precession of the ferromagnetically coupled Mn spins in (Ga,Mn)As, which is observed even with no external magnetic field applied. We concentrate on various experimental aspects of the time-resolved magneto-optical Kerr effect (TR-MOKE) technique that can be used to clarify the origin of the detected signals. We show that the measured data typically consist of several different contributions, among which only the oscillatory signal is directly connected with the ferromagnetic order in the sample.**

## I. Introduction

Diluted magnetic semiconductors, with (Ga,Mn)As as a typical model material, have attracted a significant attention for last few years due to the carrier-mediated origin of the ferromagnetism [1]. This special feature makes it possible to control magnetization of the material also by changing the concentration of the carriers [2], [3], which is very important property for future applications in semiconductor spintronics. The carrier population can be modified by different methods, among which the optical injection is the simplest one [3]. Moreover, the magnetization can be manipulated on the picosecond time scales using light pulses from ultrafast lasers [4], as illustrated by the ultrafast quenching and enhancement of ferromagnetism in GaMnAs [5]. Photoexcitation of a magnetic system can also strongly alter the equilibrium between the itinerant carriers (holes in particular), the localized spins (Mn ions), and the lattice. This in turn triggers a variety of dynamical processes in the material, which are usually investigated by the time-resolved magneto-optical Kerr effect (TR-MOKE) technique [4]. In particular, the laser-induced magnetization precession in ferromagnetic GaMnAs has been reported recently [6]-[9]. Even though TR-MOKE is a unique technique to probe the ultrafast dynamics of magnetization, the extraction of the pure magnetization dynamics from the measured TR-MOKE signals is sometimes a challenging experimental problem because a photoinduced change of the refractive index also contributes to the pump-induced MOKE signal [10], [11]. In this paper we concentrate on the identification of the part of the measured TR-MOKE signal that is directly connected with the laser-induced precession of the ferromagnetically coupled Mn spins.

## II. Experimental

The experiments were performed on an as-grown 500 nm thick ferromagnetic $Ga_{1-x}Mn_xAs$ film grown by the low temperature molecular beam epitaxy (LT-MBE) on a GaAs(001) substrate. Mn content, Curie temperature, and equilibrium hole density of the sample are $x = 0.058$, $T_C \approx 60$ K, and $p \approx 1.5 \times 10^{20}$ $cm^{-3}$, respectively. The magnetic easy axes of the sample are in-plane [8]. The sample was mounted in a cryostat, which was placed between the poles of an electromagnet, and cooled with a magnetic field $\mu_0 H = 30$ mT applied in the sample plane along the [010] crystallographic direction. The actual experiment was performed with *no external magnetic field applied* ($\mu_0 H < 0.01$ mT). The photoinduced magnetization dynamics was studied by the TR-MOKE technique [4] using a titanium sapphire laser, which generates 80 fs laser pulses with a repetition rate of 82 MHz. We used pump and probe pulses with the same wavelength. The energy fluence of the pump pulses was 7 $\mu J.cm^{-2}$ (unless noted differently in the text) and the probe pulses were always at least 10 times weaker. The polarization of the pump pulses was circular (with the hellicity controlled by a quarter-wave plate) and the probe pulses were linearly polarized along the [010] crystallographic direction. The rotation angle of the polarization plane of the reflected probe pulses was obtained by taking the *difference* of signals measured by detectors in an optical bridge detection system (for the measurements of the ellipticity change the optical bridge was equipped with an additional quarter-wave plate) [4]. Simultaneously, we measured also the *sum* of signals from the detectors, which corresponded to a probe intensity change due to the pump induced modification of the sample reflectivity. Our previous experiments revealed that the light-induced precession of magnetization is apparent as an oscillatory signal only in the polarization-independent part of the signal, which was defined as an average of the signals detected for pump pulses with the opposite circular (linear) polarizations [8]. Consequently, unless noted differently in the text, all displayed dynamics correspond to this polarization-independent part of the signal.

## III. Results and Discussion

Magneto-optics deals with phenomena induced by interaction between light and a matter exposed to a magnetic field (external or internal) [12]. The phenomena connected with a reflection of light from the surface of the sample are generally referred to as the magneto-optical Kerr effect (MOKE). The Kerr rotation $\theta$ and ellipticity $\eta$ can be expressed in the following form [11]:

$$\theta = f_\theta \cdot M, \quad (1a)$$
$$\eta = f_\eta \cdot M, \quad (1b)$$

where $M$ is the magnetization, $f_\theta$ and $f_\eta$ are functions that depend on the electronic properties of the material and that can be expressed in terms of the refractive index and the absorption coefficient. Correspondingly, the light-induced change in $\theta$ and $\eta$ consists of two components

$$\Delta\theta(t) \approx f_\theta \cdot \Delta M(t) + \Delta f_\theta(t) \cdot M, \quad (2a)$$
$$\Delta\eta(t) \approx f_\eta \cdot \Delta M(t) + \Delta f_\eta(t) \cdot M, \quad (2b)$$

where only the first term reflects directly the dynamics of the magnetization. Consequently, the dynamics of both the Kerr rotation and ellipticity should be measured and compared for the correct interpretation of the TR-MOKE data [10], [11].

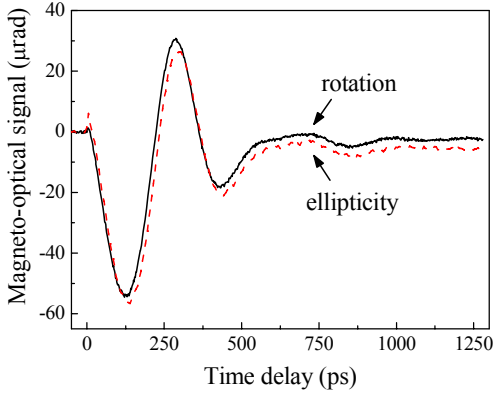

Fig. 1. Dynamics of photoinduced Kerr rotation and ellipticity measured at 1.64 eV; sample temperature 7 K.

In Fig. 1 we present typical dynamics of Kerr rotation and ellipticity measured at 7 K. The close similarity of both curves clearly shows that the measured laser-induced change in $\theta$ and $\eta$ is dominated by the first term in (2a) and (2b), respectively. It implies that the data directly reflect the dynamics of the light-induced precession of magnetization in the sample. The origin of the laser-induced magnetization precession is not clear at the moment. It is generally accepted that the oscillations are induced by a change in the magnetic anisotropy of the sample but the exact physical mechanism responsible for this anisotropy change is still discussed [6]-[9]. The measured signal contains in principle contributions from the polar Kerr rotation (PKR), which reflects the out-of-plane motion of $M$, and magnetic linear dichroism (MLD), which reflects the in-plane motion of $M$ [9]. These two contributions can be distinguished by their spectral and polarization dependence - the MLD depends strongly on both the laser wavelength and the orientation of the probe polarization (with respect to the position of $M$ in the sample plane) while for the PKR these dependencies are much weaker [9], [13]. In Fig. 2 we show the dynamics of Kerr rotation angles (KR) measured at various photon energies (laser wavelengths) - the oscillatory signal is clearly apparent at all investigated photon energies. Moreover, closer inspection of the data revealed that the oscillations are in fact a mixture of two different frequencies. These two frequencies were not apparent in the data reported previously for the same sample due to the different cooling procedure and/or laser intensity applied in [8]. We also note that only one frequency was identified in the laser-induced magnetization precession reported by other groups so far [6], [7], [9].

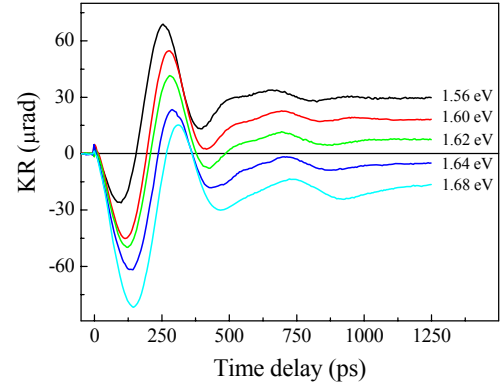

Fig. 2. Spectral dependence of dynamics of Kerr rotation angles (KR); sample temperature 7 K.

The measured dynamics of KR can be fitted well by an exponentially damped sine harmonic oscillations superimposed on a pulse-like function [8]:

$$KR(t) = \sum_{i=1}^{2} A_i \exp(-t/\tau_{Di}) \sin(\omega_i t + \varphi_i) \\ + B\left[1 - \exp(-t/\tau_1)\right]\exp(-t/\tau_2). \quad (3)$$

The oscillatory parts of the KR signal are characterized by the amplitude ($A_i$), damping time ($\tau_{Di}$), angular frequency ($\omega_i$), and phase ($\varphi_i$). The pulse-like part of the KR signal is described by the amplitude ($B$), rise time ($\tau_1$), and decay time ($\tau_2$). In Fig. 3 (a) and (b) we show two examples of the fitted curves – the points are the measured data and the solid lines are fits. The contributions of the individual terms in (3) to the measured data are compared in Fig. 3 (c). The most prominent effect for the oscillatory part of the KR signal is a phase shift between the oscillations measured at different spectral positions. For the pulse-like part of the KR signal the photon energy has an influence on the rise time and on the sign of this contribution.

The results of the analysis of the measured spectral dependence are summarized in Fig. 4. In Fig. 4 (a) we show the initial values of the reflectivity change $\Delta R/R$ that were induced in the sample by pump pulses. This signal, which monitors the initial change of the complex index of refraction of the sample, shows a sign change around the semiconductor band gap that is a feature typical for defective GaAs and (Ga,Mn)As [14], [15]. In Fig. 4 (b) we show the spectral dependence of $A_1$, $A_2$ and $B$. The magnitude of $A_1$ increases with photon energy, $A_2$ is nearly constant, and $B$ changes not only the magnitude but also the sign. The spectral dependence of $\varphi_1$ and $\varphi_2$ is shown in Fig. 4 (c). The data indicate that there is a systematic decrease of the initial phases with the photon energy (for about 70° in the investigated spectral range) but the oscillations have a constant mutual initial phase shift ($\varphi_1 - \varphi_2 = 180 \pm 10°$). The oscillation frequencies do not depend on the photon energy ($\omega_1 = 16.7 \pm 0.5$ GHz, $\omega_2 = 27.8 \pm 0.5$ GHz) and also the damping of the oscillations is similar at all photon energies (both oscillatory modes can be fitted with $\tau_D = 210 \pm 30$ ps). The rise and decay times of the pulse-like part of the KR signal do not show any systematic dependence on the photon energy. The measured spectral dependences of $A_1$ and $A_2$ are considerably weaker than those expected for MLD around the semiconductor band gap energy [13]. The magnitudes of the oscillations also do not depend strongly on the orientation of the probe polarization [19]. Consequently, both these facts strongly suggest the dominant role of PKR in the measured signal.

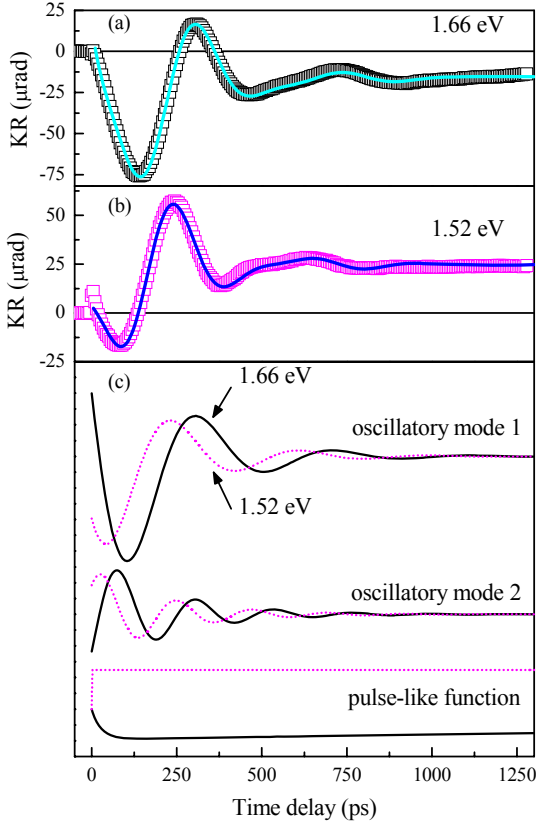

Fig. 3. Fitting procedure applied to the dynamics of KR. (a), (b) The data measured at two different photon energies (points) are fitted by (3) (solid lines). (c) The contributions of individual terms in (3) to the measured data. Sample temperature 7 K.

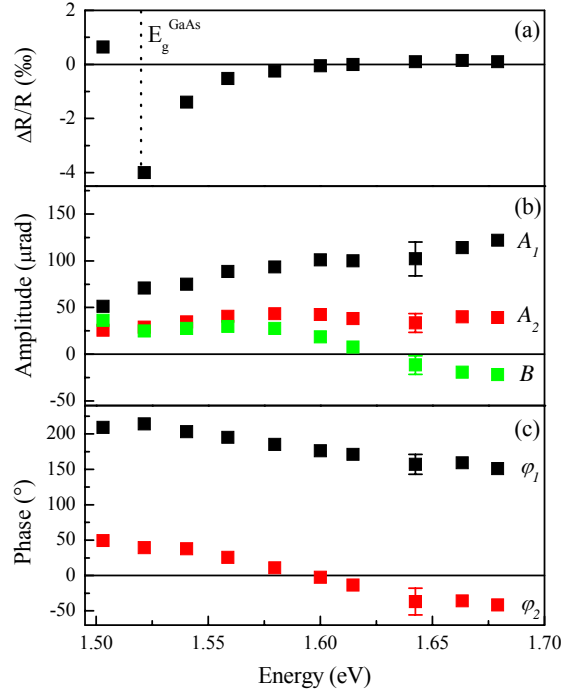

Fig. 4. Spectral dependence of the measured signals at 7 K. (a) The values of the reflectivity change $\Delta R/R$ measured in the sample for a zero time delay between pump and probe pulses, the dotted line represents the band gap energy of GaAs. (b) Spectral dependence of $A_1$, $A_2$ and $B$. (c) Spectral dependence of $\varphi_1$ and $\varphi_2$.

In Fig. 5 (a) we show the dynamics of KR measured for the right ($\sigma^-$) and left ($\sigma^+$) circularly polarized pump pulses at $T$ = 80 K, which is 20 K *above* the Curie temperature $T_C$ in the investigated sample. In Fig. 5 (b) we show the polarization-independent part and the polarization-dependent part of the measured KR signal. The absence of oscillations in the polarization-independent part of KR for $T > T_C$ is in accord with their assignment to the laser-induced precession of the ferromagnetically coupled Mn spins (that is obviously absent above $T_C$). On the other hand, the sizable magnitude of the pulse-like function in the polarization-independent part of the measured KR signal and the presence of the polarization-dependent part of the measured KR signal for $T > T_C$ clearly indicate that these two contributions (or at least their parts) *are not directly connected with the ferromagnetic order* in (Ga,Mn)As. The origin of these contributions is not clear at the moment because the magneto-optical response of (Ga,Mn)As is rather complicated problem. In particular, the *sp-d* interaction in Mn-doped semiconductors is responsible for a strong magneto-optical signal even in the paramagnetic phase [16]. Also the spin-polarized electrons, which are photoinjected by absorption of circularly polarized light [17], contribute significantly to the polarization-dependent part of KR [18].

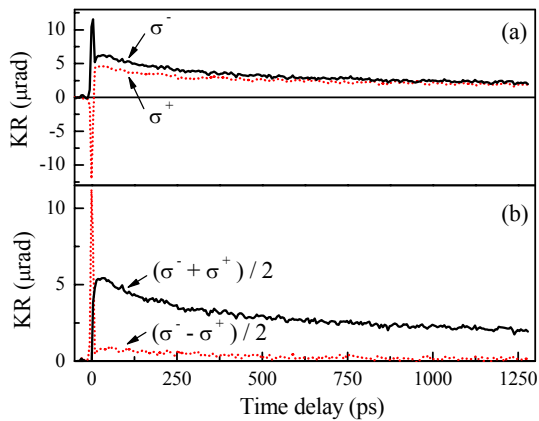

Fig. 5. Dynamics of KR measured at 80 K (that is 20 K above $T_C$); photon energy 1.64 eV. (a) KR measured for $\sigma^-$ and $\sigma^+$ circularly polarized pump pulses; (b) Polarization-independent part $(\sigma^- + \sigma^+)/2$ and polarization-dependent part $(\sigma^- - \sigma^+)/2$ of KR signal. The data were measured at 10 times higher pump intensity (70 µJ.cm$^{-2}$) than that used at 7 K.

## IV. CONCLUSION

We studied the laser-induced precession of the ferromagnetically coupled Mn spins in (Ga,Mn)As by the time-resolved magneto-optical Kerr effect. We showed that the oscillatory signal is present both in the Kerr rotation and ellipticity measurements that proves its "magnetic" origin. Our data indicated that only the oscillatory signal is directly connected with the ferromagnetic order in the sample and that the contribution from the polar Kerr rotation should dominante in the measured signal. We also identified two distinct oscillation components in the measured signal (with a mutual phase shift of 180°). This fact shows that the laser-induced precession of magnetization can be possibly used as an efficient diagnostic method for the investigation of ferromagnetic semiconductors.

ACKNOWLEDGMENT

This work was supported in part by Ministry of Education of the Czech Republic in the framework of the research centre LC510, the research plans MSM0021620834 and AV0Z1010052, by the Grant Agency of the Charles University in Prague under Grant No. 252445, and by the Grant Agency of Academy of Sciences of the Czech Republic Grants FON/06/E001, FON/06/E002, and KAN400100652.

REFERENCES

[1] T. Jungwirth, J. Sinova, J. Mašek, A. H. MacDonald, "Theory of ferromagnetic (III,Mn)V semiconductors," Rev. Mod. Phys. 78, pp. 809-864, 2006.
[2] H. Ohno, D. Chiba, F. Matsukura, T. Omiya, E. Abe, T. Dietl, Y. Ohno, K. Ohtani, "Electric-field control of ferromagnetism," Nature 408, pp. 944-946, 2000.
[3] S. Koshibara, A. Oiwa, M. Hirasawa, S. Katsumoto, Y. Iye, C. Urano, H. Takagi, H. Munekata, "Ferromagnetic order induced by photogenerated carriers in magnetic III-IV semiconductor heterostructures of (In,Mn)As-GaSb," Phys. Rev. Lett. 78, 4617-4620, 1997.
[4] J. Wang, Ch. Sun, Y. Hashimoto, J. Kono, G.A. Khodaparast, L. Cywinski, L.J. Sham, G.D. Sanders, Ch.J. Stanton, H. Munekata, "Ultrafast magneto-optics in ferromagnetic III-V semiconductors," J. Phys.: Condens. Matter 18, pp. R501-R530, 2006.
[5] J. Wang, I. Cotoros, K.M. Dani, X. Liu, J.K. Furdyna, D.S. Chemla, "Ultrafast enhancement of ferromagnetism via photoexcited holes in GaMnAs," Phys. Rev. Lett. 98, 217401, 2007.
[6] A. Oiwa, H. Takechi, H. Munekata, "Photoinduced magnetization rotation and precessional motion of magnetization in ferromagnetic (Ga,Mn)As," J. Supercond. 18, pp. 9-13, 2005.
[7] J. Qi, Y. Xu, N. H. Tolk, X. Liu, J. K. Furdyna, I. E. Perakis, "Coherent magnetization precession in GaMnAs induced by ultrafast optical excitation," Appl. Phys. Lett. 91, 112506, 2007.
[8] E. Rozkotová, P. Němec, P. Horodyská, D. Sprinzl, F. Trojánek, P. Malý, V. Novák, K. Olejník, M. Cukr, T. Jungwirth, submitted for publication to Appl. Phys. Lett., preprint arXiv:0802.2043
[9] Y. Hashimoto, S. Kobayashi, H. Munekata, "Photoinduced precession of magnetization in ferromagnetic (Ga,Mn)As," Phys. Rev. Lett. 100, 067202, 2008.
[10] B. Koopmans, M. van Kampen, J. T. Kohlhepp, W. J. M. de Jonge, "Ultrafast magneto-optics in nikel: magnetism or optics?," Phys. Rev. Lett. 85, pp. 844-847, 2000.
[11] E. Kojima, R. Shimano, Y. Hashimoto, S. Katsumoto, Y. Iye, M. Kuwata-Gonokami, "Observation of the spin-charge thermal isolation of ferromagnetic Ga$_{0.94}$Mn$_{0.06}$As by time-resolved magneto-optical measurements," Phys. Rev. B, 193203, 2003.
[12] A. K. Zvezdin, V. A. Kotov, Modern Magnetooptics and Magnetooptical Materials, Institute of Physics Publishing, Bristol and Philadelphia, 1997, pp 1-62.
[13] A. V. Kimel, G. V. Astakhov, A. Kirilyuk, G. M. Schott, G. Karczewski, W. Ossau, G. Schmidt, L. W. Molenkamp, Th. Rasing, "Observation of giant magnetic linear dichroism in (Ga,Mn)As," Phys. Rev. Lett. 94, 227203, 2005.
[14] S Janz, U. G. Akano, I. V. Mitchell, "Nonlinear optical response of As+-ion implanted GaAs studied using time resolved reflectivity," Appl. Phys. Lett. 68, pp. 3287-3289 (1996).
[15] S. Kim, E. Oh, J. U. Lee, D. S. Kim, S. Lee, J. K. Furdyna, "Effect of point defect and Mn concentration in time-resolved differential reflection in GaMnAs," Appl. Phys. Lett. 88, 262101, 2006.
[16] D. U. Bartholomev, J. K. Furdyna, A. K. Ramdas, "Interband Faraday rotation in diluted magnetic semiconductors: Zn$_{1-x}$Mn$_x$Te and Cd$_{1-x}$Mn$_x$Te," Phys. Rev. B 34, pp. 6943-6950, 1986.
[17] R. D. R. Bhat, P. Němec, Y. Kerachian, H. M. van Driel, J. E. Sipe, A. L. Smirl, "Two-photon spin injection in semiconductors," Phys. Rev. B 71, 035209, 2005.
[18] A. V. Kimel, G. V. Astakhov, G. M. Schott, A. Kirilyuk, D. R. Yakovlev, G. Karczewski, W. Ossau, G. Schmidt, L. W. Molenkamp, Th. Rasing, "Picosecond dynamics of the photoinduced spin polarization in epitaxial (Ga,Mn)As films," Phys. Rev. Lett. 92, 237203, 2004.
[19] E. Rozkotová et al., unpublished.

Corresponding author: P. Němec (e-mail: nemec@karlov.mff.cuni.cz).